\documentclass[aip, jcp, floatfix, amsfonts, superscriptaddress, reprint]{revtex4-1}

\usepackage{times, gensymb, amsmath, amsthm, amssymb, bm, graphicx, setspace, siunitx, graphicx} \pdfoutput=1

\usepackage[normalem]{ulem}
\usepackage[]{color}
\usepackage[usenames,dvipsnames,svgnames]{xcolor}
\usepackage{soul}

\usepackage[stretch=10]{microtype}

\usepackage{hyperref}
\hypersetup{colorlinks=true, linkcolor=blue, citecolor=blue}

\newcommand{\ket}[1]{|#1\rangle}

\newcommand{\ketbra}[2]{| #1 \rangle \langle #2 |}

\newcommand{\Lket}[1]{|#1\rangle\!\rangle}
\newcommand{\Lbra}[1]{\langle\!\langle#1|}
\newcommand{\Lbraket}[2]{\langle\!\langle #1 | #2 \rangle\!\rangle}

\newcommand{\stretchLbraket}[2]{\left\langle\!\middle\langle #1 \middle| #2 \middle\rangle\!\right\rangle}

\newcommand{\iso}[1]{\langle#1\rangle_\text{iso}}
\newcommand{\stretchiso}[1]{\left\langle#1\right\rangle_\text{iso}}
\renewcommand{\vec}[1]{\bm{#1}}

\DeclareMathOperator{\tr}{Tr}
\DeclareMathOperator*{\argmin}{argmin}

\newcommand{\muga}{\mu_{g\alpha}}
\newcommand{\mugb}{\mu_{g\beta}}
\newcommand{\muaf}{\mu_{\alpha f}}
\newcommand{\mubf}{\mu_{\beta f}}

\newcommand{\rhopp}{\rho_\text{PP}^{(2)}}
\newcommand{\Rpp}{R_\text{PP}}

\begin{document}

\title{Inverting pump-probe spectroscopy for state tomography of excitonic systems}
\date{\today}
\author{Stephan Hoyer}
\affiliation{Department of Physics, University of California, Berkeley, CA 94720, USA}
\affiliation{Berkeley Quantum Information and Computation Center, University of California, Berkeley, CA 94720, USA}
\author{K. Birgitta Whaley}
\affiliation{Berkeley Quantum Information and Computation Center, University of California, Berkeley, CA 94720, USA}
\affiliation{Department of Chemistry, University of California, Berkeley, CA 94720, USA}

\begin{abstract}
We propose a two-step protocol for inverting ultrafast spectroscopy experiments on a molecular aggregate to extract the time-evolution of the excited state density matrix.
The first step is a deconvolution of the experimental signal to determine a pump-dependent response function. The second step inverts the quantum state of the system from this response function, given a model for how the system evolves following the probe interaction.
We demonstrate this inversion analytically and numerically for a dimer model system, and evaluate the feasibility of scaling it to larger molecular aggregates such as photosynthetic protein-pigment complexes. Our scheme provides a direct alternative to the approach of determining all Hamiltonian parameters and then simulating excited state dynamics.
\end{abstract}

\maketitle

\section{Introduction}

Ultrafast nonlinear spectroscopy allows us to experimentally observe excited state dynamics in molecular aggregates, and in particular, energy transfer essential to the function of natural light harvesting systems \cite{Mukamel1995, Blankenship2002, May2011}.
The existence of these experimental tools prompts a natural question: is it possible to use spectroscopic measurements to directly infer the excited state of such systems?
A complete answer to this question would be a procedure for quantum state tomography (QST), that is, for reconstruction of the full density matrix describing the quantum state \cite{Nielsen2000, Thew2002}. State tomography is a technique that has found widespread application for validating and characterizing quantum devices designed as components for quantum computation.
Such full characterization of an exciton state over multiple pigments, beyond a mere classical probability distribution, would offer information essential to understanding the explicitly quantum features of energy transport, which include coherence \cite{Engel:2007vm}, entanglement \cite{Sarovar2009} and possibly other types of non-trivial quantum dynamics \cite{Hoyer2010, Hoyer:2011fg, Bradler:2010up, Nalbach:2011ef}.
In this work, we show that under appropriate conditions and assumptions, QST of excited states can be performed from the results of a series of pump-probe type ultrafast spectroscopies.

The most sophisticated non-linear technique for resolving energy transfer dynamics is the two-dimensional (2D) photon-echo, in which the time delays between three ultrafast pulses are manipulated to provide a 2D map between pump and probe frequencies at fixed time delays \cite{Cheng2009,Abramavicius2009}.
These two-dimensional maps allow for direct visualization of the relationship between excitation and emission energies as a function of delay time.
More formally, 2D spectroscopy is usually interpreted in the limit of impulsive interactions. In this approximation, it provides snapshots of the 3rd-order response function \cite{Mukamel1995}. Important applications of 2D spectroscopy to photosynthetic systems have included resolving energy transfer pathways \cite{Brixner2005} and the dynamics of electronic quantum beats \cite{Engel:2007vm, SchlauCohen:2012wh, Panitchayangkoon2011, Collini2010}.
In contrast, pump-probe spectroscopy (also known as transient absorption) is a simpler type of 3rd-order spectroscopy that historically predates 2D. In a pump-probe setup, a pump pulse excites the system, which is probed at some time later by a probe pulse.
Because of its relative ease of experimental implementation, pump-probe was used to follow ultrafast energy transfer dynamics in photosynthesis long before 2D spectroscopy. For example, it provided the first evidence of electronic quantum beats in photosynthetic pigment-protein complexes, in 1997 \cite{Savikhin1997}.
Pump-probe provides less information than the 2D photon echo, because the pump-probe signal can be obtained by integrating over the excitation axis in a 2D spectra \cite{SchlauCohen:2012du}.
However, for the purposes of this work, pump-probe has a clear advantage, namely, that it can be directly interpreted as a measurement of the state created by the pump pulse.
Formally, the pump dependence is entirely contained within the change in the density matrix of the system after interacting with the pump \cite{Jonas:1995wy}.

In the past, time-resolved spectra such as pump-probe have been analyzed by simultaneously or concurrently fitting spectral components, known as decay associated or species associated spectra, with a kinetic model \cite{vanStokkum:2004tv, Wohlleben:2003vn}.
These techniques are powerful, as evidenced by their widespread application to experiments. However, much of the kinetic information they reveal can be seen more directly in 2D spectra.
Moreover, kinetic models, although adequate for many purposes, cannot describe more complex dynamics, such as those deriving from quantum beats or from a non-Markovian bath. Our approach to QST side-steps the issues of extending such integrated analyses by focusing on identifying the quantum state directly.

Recently, it was shown that a combination of photon-echo measurements of excitonic systems can be combined to perform quantum process tomography of excitonic dimers, either by using differently colored pulses  \cite{Yuen-Zhou2011} or by combining peak amplitudes from a set of 2D spectra \cite{Yuen-Zhou2011a}.
Process tomography \cite{Nielsen2000, Mohseni2008a} is more general than state tomography, because it specifies the full set of possible quantum evolutions for a system given any initial condition.
This makes it well suited to characterizing gates for quantum computation, which are by definition designed to handle any possible input state.
However, determining the full process matrix is expensive: it requires inverting at least $d^4 - d^2$ real parameters for a $d$-dimensional Hilbert space, in contrast to $d^2$ parameters for state tomography. Moreover, for analysis of complex molecular dynamics in condensed phases, such information,
although potentially helpful, is not necessary, because most trajectories contained in the process matrix start from initial conditions that are implausible for a molecular aggregate. Indeed, typical theoretical investigations of dynamics in light harvesting systems \cite{Ishizaki2009a, Shim:2012el} follow dynamics after excitation for only a limited set of plausible initial states, such as the states which absorb sunlight or excitations from neighboring antenna complexes.
Finally, the relative simplicity of state tomography helps to simplify consideration of new theoretical approaches to tomography, particularly because process tomography is often based on as series of state tomographies \cite{Mohseni2008a}.

In this paper, we present a new approach to state tomography of excitonic systems based on pump-probe spectroscopy. Our approach is based on a two stage protocol that separates the easy (field based) and hard (system based) parts of the inversion process.
This yields several advantages over prior approaches, including the ability to use arbitrarily shaped laser pulses and to perform the first inversion even when the second inversion is not possible.
After presenting the details of each of these inversions, we demonstrate their feasibility by applying them to invert the simulated spectra of a model dimer with a Markovian environment. We close with a consideration of the conditions under which state inversion would be feasible for a natural light-harvesting system, the FMO complex of green sulfur bacteria.

\section{Recipe for pump-probe spectroscopy}
\label{sec:recipe}

We begin by presenting the specific theoretical formalism for pump-probe spectroscopy that we propose to invert.
The measured signal in any 3rd-order spectroscopy experiment, including pump-probe, is a function of the 3rd-order polarization \cite{Mukamel1995}. This 3rd-order polarization depends on three interactions between the applied fields and the sample, with the time-ordering of these interactions enforced by time delays of the pulses and by looking at the signal emitted in a particular phase-matched direction. For a pump-probe experiment, the first two interactions happen with the same pulse, the pump, and the last interaction is with the probe pulse. The phase-matched condition is that the signal is observed in the direction of the probe.
Based on this phase-matched geometry and the response function formalism \cite{Abramavicius2009}
(see Appendix \ref{sec:response_function}), we can combine the allowed time orderings to write the nonlinear polarization for a pump-probe experiment under the rotating wave approximation as
\begin{align}
	P^{(3)}(t) =& \int_0^\infty dt_3 \Rpp (t_3, \rhopp (t-t_3) ) E_\text{pr}^+ (t - t_3),
	\label{eq:pp-polarization}
\end{align}
in terms of the pump-probe response
\begin{align}
	\Rpp(t, \rhopp) &= \frac{i}{\hbar} \tr\left[\mu^{(-)} G(t) V^{(+)} \rhopp \right].
	\label{eq:pp-response}
\end{align}
This pump-probe response is of identical form to the linear response function \cite{Mukamel1995}, but with the electronic ground-state density matrix $\rho_0$ replaced by the second-order contribution to the density matrix $\rhopp$ that contributes to the phase-matched signal observed in a pump-probe experiment (that is, with signal wave-vector $\vec{k}_\text{S}=\vec{k}_\text{pr}$).
The quantities ${E}^+_\text{pu}(t)$ and $E^+_\text{pr}(t)$ denote the complex envelopes of the pump and probe pulses, respectively, with $E^{-} \equiv (E^{+})^\ast$. The dipole operators $\mu^{(-)} = \sum_n d_n a_n$ and $\mu^{(+)} = {(\mu^{(-)})}^\dagger$, with $a_n$ as the annihilation operator for an electronic excitation on pigment $n$ and $d_n$ the corresponding dipole moments. The Liouville space operator $G(t)$ is the retarded material Green function for evolution for time $t$ and $V^{(\pm)}\ \cdot \equiv [\mu^{(\pm)}, \cdot]$. Formally, the portion of the second order contribution to the density matrix which contributes to the signal is given by
\begin{align}
	\rhopp(t) = \left(\frac{i}{\hbar}\right)^2 2 \sum_{\pm} &\iint_0^\infty  dt_2 dt_1  G(t_2) V^{(\pm)} G(t_1) V^{(\mp)} \rho_0 \notag \\[-3pt]
	&\times E_\text{pu}^\pm (t - t_2) E_\text{pu}^\mp (t - t_2 - t_1).
	\label{eq:rho2}
\end{align}
In deriving Eqs.~(\ref{eq:pp-polarization}--\ref{eq:rho2}), we employed the rotating wave approximation (accurate for resonant excitation \cite{Abramavicius2009}) and neglected those terms from the double-quantum-coherence contribution ($\vec{k}_\text{S} \neq \vec{k}_\text{pr}$).
Accordingly, we can safely neglect the possibility of multiple excitations in the calculation of $\rhopp$. In Appendix \ref{sec:properties_of_rho_2} we prove that the excited state portion of $\rhopp$ is both equal to the excited state portion of the full density matrix and is itself a valid (but unnormalized) density matrix.

The core of our proposed quantum state tomography is the sequential inversion of Eqs.~(\ref{eq:pp-polarization}--\ref{eq:rho2}).
The remainder of this section discusses additional details relevant to simulating experimental signals to test our inversion procedure.
We emphasize that these expressions hold under the very general conditions, requiring only the rotating wave approximation, that all applied fields are weak and negligible overlap between pump and probe pulses. No assumptions were made concerning the shapes of these pump and probe pulses.
Our decomposition here is similar to the window-doorway picture for the pump-probe signal \cite{Mukamel1995}, but here we have separated out the influence of the control fields, even when not in the impulsive ``snapshot'' limit.
Related expressions in terms of a convolution of pump and probe components have been shown to facilitate analysis of pump-probe experiments with shaped probes \cite{Polli:2010hr}.

\subsection{Detection scheme and probe convolution}
\label{sub:detection_scheme}

In a typical pump-probe experiment, the probe pulse has a fixed time-envelope, subject to a variable delay time $T$ between the two pulses. Accordingly, we may substitute $E^+_\text{pr}(t) = E_\text{pr} (t - T)$.
Likewise, experimental signals are most directly interpreted in the frequency domain, so we now consider the Fourier transform of the nonlinear polarization, $P^{(3)}(\omega) = \int dt \ e^{i\omega(t-T)} P^{(3)}(t)$, calculated
relative to the probe delay $T$. We can also write the pump-probe response in the Fourier domain, $\Rpp(\omega,\rhopp) = \int dt \ e^{i\omega t} \Rpp(t,\rhopp)$.
In terms of these quantities in the frequency domain with the explicit probe delay $T$, we can then replace Eq.~\eqref{eq:pp-polarization} with a one-dimensional convolution,
\begin{align}
	P^{(3)}(\omega, T)=& \int_{-\infty}^\infty d\tau\ \Rpp(\omega,\rhopp(\tau)) E_\text{pr} (\tau - T) e^{i\omega(\tau-T)}.
	\label{eq:convolution}
\end{align}
To obtain this relation, we substituted $t_3 = t - \tau$ and extended the lower limit of the integral in Eq.~\eqref{eq:pp-polarization} to $-\infty$, because by definition $G(t) = 0$ for $t < 0$.
We cannot simply turn this convolution into a multiplication by taking the Fourier transform of these quantities with respect to $T$, because for small or negative delay times $T$, there are contributions from signals where the pump does not necessarily interact before the probe.
If $\Rpp(\omega,\rhopp(\tau))$ does not vary appreciably over the duration of the probe pulse, then the equation above is the Fourier transform of the probe field envelope, so we can approximate
\begin{align}
	P^{(3)}(\omega, T) &\approx E_\text{pr}(\omega) \Rpp(\omega,\rhopp(T)).
	\label{eq:naive-inversion}
\end{align}
In the limit of a completely impulsive probe, $E_\text{pr}(t) \approx E_0 \delta(t)$ and thus $E_\text{pr}(\omega)$ is constant, so the nonlinear polarization and the pump-probe response are equal up to a constant of proportionality.

We measure the non-linear polarization $P^{(3)}(t)$ by detecting the corresponding signal field $E_\text{S}(t) \propto i P^{(3)}(t)$ \cite{Mukamel1995}. Here we consider heterodyne detection, either with the probe pulse as in a standard ``self-heterodyned'' pump-probe setup, or with a separate local-oscillator (LO) pulse. The use of a separate local-oscillator is possible in a transient-grating setup, in which the pump pulse is replaced by two otherwise identical pumps with different wavevectors
$\vec k_1$ and $\vec k_2$, so that the signal wavevector $\vec k_\text{S} = -\vec k_1 + \vec k_2 + \vec k_3$ does not match the probe wavevector $\vec k_3$. Mathematically, this transient-grating signal yields the same non-linear polarization as in pump-probe, although it raises experimental complications by requiring phase-stability with an additional pulse. In heterodyne detection, the absolute value squared of the sum of the signal and local-oscillator (or probe) fields can be spectrally dispersed and measured in the frequency domain \cite{Schlau-Cohen2011}. Typically, the signal field is much smaller than than of the local oscillator, so upon subtracting away the local oscillator contribution, the measured signal $S(\omega)$ is proportional to $\text{Re}[E_\text{S}(\omega) E_\text{LO}^\ast(\omega)]$, and thus
\begin{align}
	S(\omega,T) \propto \text{Im}[{P^{(3)}}(\omega,T) E^\ast_\text{LO}(\omega)].
	\label{eq:heterodyne-detect}
\end{align}
This equation is a multiplication in the frequency domain. Hence it is a convolution, and takes on similar form to Eq.~\eqref{eq:convolution} when expressed in the time-domain.
In the pump-probe setup, $E_\text{LO} = E_\text{pr}$, so the signal for a fast probe given by inserting Eq.~\eqref{eq:naive-inversion} yields
\begin{align}
	S(\omega, T) &\propto |E_\text{pr}(\omega)|^2 \text{Im} \Rpp(\omega,\rhopp(T)).
\end{align}
In this case, the signal only depends on the imaginary (absorptive) part of the non-linear polarization $P^{(3)}$ and the pump-probe response. In the alternative transient grating setup, as long as one is still in the limit of a fast probe, applying a $\pi/2$ phase shift to the now distinct local oscillator pulse allows for obtaining the real (dispersive) part of the pump-probe response function in a similarly direct manner  \cite{Schlau-Cohen2011}.
More generally, heterodyne detection with and without a $\pi/2$ phase shift allows for obtaining both real and imaginary parts of the non-linear polarization, respectively.

\subsection{Pump-probe response function}
\label{sub:response_function}

To isolate the effect of the probe, the pump-probe response function given by Eq.~\eqref{eq:pp-response} can be written as
\begin{align}
	\Rpp(t,\rhopp)&= \tr\left[ \mathcal{P}(t) \rhopp \right],
	\label{eq:pp-response-measure}
\end{align}
with the pump-probe response operator $\mathcal{P}(t)$ defined as
\begin{align}
	\mathcal{P}(t) &= \frac{i}{\hbar} \mu^{(-)} G(t) V^{(+)} = \frac{i}{\hbar}[\mu^{(-)}(t), \mu^{(+)}(0)],
	\label{eq:pp-response-operator}
\end{align}
where $\mu^{(\pm)}(t)$ denotes $\mu^{(\pm)}$ in the Heisenberg picture. This is similar but not equivalent to a family of quantum measurements \cite{Nielsen2000} parametrized by the continuous time variable $t$ (or frequency $\omega$ in the Fourier domain), since $\Rpp$ can be complex valued.
Accordingly, the pump-probe response can be interpreted as the projection of $\rhopp$ onto $\mathcal{P}(t)$, where these are viewed as
vectors in Liouville space \cite{Mukamel1995},
\begin{align}
	\Rpp(t,\rhopp) = \stretchLbraket{\mathcal{P}(t)}{\rhopp}.
	\label{eq:pp-response-vec}
\end{align}
Individual components of the pump-probe response operator $\Lbraket{\mathcal{P}(\omega)}{\alpha}$ are equivalent to the species associated spectra of the state $\Lket{\alpha}$ \cite{vanStokkum:2004tv}.

In most spectroscopy experiments, the signal is an ensemble measurement summed over all possible molecular orientations and static disorder of Hamiltonian parameters (inhomogeneous broadening). Accordingly, the pump-probe response in Eq.~\eqref{eq:pp-response} should be replaced by its average over molecular orientations and static disorder. The orientational average can be handled elegantly using the expression for the pump-probe response in Eq.~\eqref{eq:pp-response-measure}: in the magic angle $\theta \approx 54.7\degree$ (MA) relative polarization configuration between the pump and probe pulses \cite{Hochstrasser2001}, the quantities $\mathcal{P}(t)$ and $\rhopp$ can simply be replaced by their independent isotropic averages,
\begin{align}
	\left\langle \Rpp(t,\rhopp) \right\rangle_\text{MA} &= \tr\left[ \Big\langle\mathcal{P}(t)\Big\rangle_\text{iso} \stretchiso{\rhopp} \right].
	\label{eq:pp-response-measure-iso}
\end{align}
By virtue of the properties of isotropically averaged tensors \cite{Craig1984}, these independent isotropic averages are equal to the average of the quantities obtained from the $xx$, $yy$ and $zz$ configurations. In contrast, the ensemble average over static disorder cannot be factorized this way in general,
because under static disorder the pump-probe operator and density matrix are correlated, and altering the system Hamiltonian (e.g., to shift transition energies) changes both quantities systematically.

\section{Inversion protocols}

\subsection{Deconvolution of the pump-probe signal}
\label{sec:inverting_the_pump_probe_response}

The first stage of our inversion protocol is a double-deconvolution to determine the complex valued pump-probe response function
$\Rpp(T,\rhopp)$ from the results of a series of heterodyne measurements, i.e., the signal $S(\omega,T)$. We need such a double-deconvolution procedure because the results of heterodyne detection depend on a (trivial) convolution over the non-linear polarization, which in turn depends on a convolution over the response function [see Eqs.~\eqref{eq:convolution} and \eqref{eq:heterodyne-detect}]. Since the excited state density matrix is entirely contained in the pump-probe response function (Appendix \ref{sec:properties_of_rho_2}), this inversion retains all information about the quantum state. However, it is not immediately clear that the real (dispersive) part of the response function contains useful information independent of the imaginary (absorptive) part, which is the portion measured by usual pump-probe experiments.

Inverting the signal to obtain the pump-probe response function is a non-trivial but important task, since, as pointed out above, the signal is directly proportional to the response only when the probe pulse is much faster than all energy transfer dynamics. Such pulses can be difficult to realize experimentally.
The need for a full inversion to obtain the response function is particularly relevant for understanding experiments which show fast oscillations due to quantum beats, whether these are of electronic, vibrational or mixed origin.
In such cases, the fast probe assumption of Eq.~\eqref{eq:naive-inversion} is not valid.
We shall refer to the use of this approximate description for inversion as the ``naive'' approach. In contrast, a proper treatment of this inverse problem would attempt to undo the convolution in Eq.~\eqref{eq:convolution}.

To address this challenge, we suggest the use of standard deconvolution techniques \cite{Hansen:2010vw} based on general-form Tikhonov regularization (also known as ridge regression), which we describe in detail in Appendices \ref{sec:tikhonov_regularization} and \ref{sec:parameter_selection_techniques}.
The response function can then be obtained from two sequential 1D deconvolutions.
First, we invert the non-linear polarization $P^{(3)}(\omega,T)$ from the measured signal $S(\omega, T)$ recorded at each choice of delay time $T$. The relationship between these signals is simple multiplication by the probe (or local oscillator) field in the frequency domain, so this step only uses the deconvolution to smooth the reconstruction along the $\omega$-axis.
Second, we invert the response function $\Rpp(\omega,\rhopp(T))$ from one-dimensional deconvolutions of the non-linear polarization $P(\omega,T)$, for each fixed value of $\omega$. For this inversion, we only use experimental data with the delay between pump and probe pulses long enough so that we can ignore pulse overlap effects.
Otherwise, we would be including non-pump-probe contributions to the signal.
However, we also reconstruct the pump-probe response at shorter times to appropriately handle boundary conditions, since the probe convolution means that these values for the response function contribute to the nonlinear polarization inside our region of interest.

This first stage in the inversion of pump-probe experiments requires only the detection results, i.e., the signal $S(\omega,T)$, and an excellent characterization of the probe and local oscillator fields. No system information is required at all. Likewise, we have sacrificed no information from our measurement about the internal system information, including its quantum state.
Thus in principle this stage can be performed with high accuracy for any system, no matter how complex its internal degrees of freedom.

\subsection{Inverting the quantum state}
\label{sec:qst_general}

The second step to complete the state tomography is to invert the pump-probe response function $\Rpp(\omega, \rhopp(T))$ to obtain the quantum state $\rhopp(T)$.
This is certainly the harder step, since it requires the ability to construct the pump-probe response of arbitrary states. The necessary information is contained in the pump-probe operator $\mathcal{P}(t)$ given by Eq.~\eqref{eq:pp-response-operator}; calculating this requires both the transition dipole moments and a model for dynamics of the 1-exciton coherences between the probe and signal interactions.
Essentially the same information is necessary to implement proposed algorithms for quantum process tomography \cite{Yuen-Zhou2011, Yuen-Zhou2011a}.
However, we emphasize that we do not need to know the nature of the initial state created by the pump pulse nor any details of the energy transfer dynamics in the 1-exciton subspace. The lack of required microscopic dynamical information is significant, since exact energy transfer dynamics are non-trivial to calculate from first principles \cite{Pachon:2012vz}.

Here we consider a simple protocol for state tomography, based on an assumed model for calculating the pump-probe response. It is by no means the only such possible state tomography protocol: we choose it because it is straightforward to implement, and turns out to be relatively robust to imperfections such as static disorder.
To perform the inversion, we propose to extract an estimate of the excited-state electronic density matrix $\hat\rho_e(\tau)$ from the estimated response function $\hat \Rpp(\omega_i, \tau)$ at that delay time $\tau$, for each frequency $\omega_i$ matching the single-exciton transition energies.
The relationship between the vector of pump-probe response measurements and the density matrix elements at any fixed time delay is linear [see Eq.~\eqref{eq:pp-response-vec}], so as long as this map is non-singular, we can solve for the density matrix by simply applying the matrix inverse to the vector formed by these estimated response function points. It is possible that in some circumstances this reconstruction will not yield a valid density matrix, since we did not include the constraint that the reconstruction be positive semi-definite. In this case, then a best estimate to minimize the mean-squared-error of the reconstruction should be obtained using techniques based on maximum likelihood \cite{James:2001ut}, although we do not encounter this issue for the examples we consider in this paper.

An additional important practical step is the choice of Liouville space in which the extracted state lies.
Our results so far hold for transition dipole operators and time evolution without any particular restrictions concerning electronic vs vibrational states or the Hamiltonian we use to describe our system. However, as a practical matter for excitonic energy transfer in light harvesting systems, we are most interested in inverting the electronic degree of freedom.
The electronic portion of $\rhopp$ has useful structure: namely, it only includes nonzero elements in two blocks, the 0- and 1-excitation subspaces.
We denote the projection of the density matrix $\rho$ onto these subspaces by $\rho_g$ and $\rho_e$, respectively.
There is only one electronic state in the 0-excitation subspace (the ground state $g$), so the electronic portion of $\rho_g$ must be in that state, $\ketbra{g}{g}$.
In the Markov limit, or for delay times much longer than the bath relaxation time, the vibrational portion of $\rho_g$ will be in thermal equilibrium  $\rho^\text{B}_\text{eq}$. These facts determine $\rho_g$, up to a constant of proportionality: the ground state population. Because total probability is conserved in the process of laser excitation, $\tr \rhopp = 0$, so the ground state population is related to the excited state population by $\tr\rho^{(2)}_g = -\tr\rho^{(2)}_e$.
Accordingly, we can write
\begin{align}
	\rhopp &= -\left(\ketbra{g}{g} \otimes \rho^\text{B}_\text{eq} \right) \tr \big[\rho^{(2)}_e\big] + \rho^{(2)}_e.
	\label{eq:rhopp-rhoe}
\end{align}
In this case, the pump-dependence in the pump-probe signal [Eq.~\eqref{eq:rho2}] is entirely contained in the excited state portion of the density matrix. Since for weak fields $\rho_e \approx \rho^{(2)}_e$,
with Eqs.~\eqref{eq:pp-response-vec} and \eqref{eq:rhopp-rhoe} we have a linear map from any excited state density matrix $\rho_e$ to the corresponding pump-probe response.
For a system with $n$ electronic states, we can parameterize this unnormalized density matrix in terms of a linear combination of $n^2$ real parameters \cite{Thew2002}, since the excited state density matrix is positive (see Appendix \ref{sec:properties_of_rho_2}) and thus Hermitian.

A brief discussion of the scalability of this approach is in order.
Based on the real and imaginary parts of the pump-probe response function in the magic angle configuration, our state tomography protocol in principle has $2n$ independent real parameters from which to extract the $n^2$ real parameters (including normalization) necessary to describe an arbitrary excited state density matrix of $n$ electronic states \cite{Thew2002}. Accordingly, we cannot necessarily expect this procedure to scale beyond a dimer ($n=2$), for which we numerically demonstrate the success and stability of this inversion procedure in the next section.
The recently proposed quantum process tomography algorithm based on peak and cross-peak amplitudes in 2D spectroscopy \cite{Yuen-Zhou2011a} has similar scaling difficulties. It requires determining $n^4-n^2$ real parameters in the process matrix from at most $12n^2$ possible experimental measurements: the real and imaginary signals, $n$ coherence and $n$ rephasing frequencies, at most 3 independent polarization configurations and 2 phase-matched geometries.
These estimates, however, hold only for this specific approach and with a randomly oriented ensemble.
Oriented or single molecule measurements offer a much larger number of independent polarization measurements, a point we will return to Section~\ref{sec:scaling_to_larger_systems}.

\section{Example: dimer model}
\label{sec:qst_for_a_dimer}

To understand in more detail how the quantum state determines the pump-probe signal, we consider the case of the signal for a dimer of coupled pigments.
In general, we can write an effective Hamiltonian for the electronic excited states of a dimer in the form
\begin{align}
	H_\text{el} = E_1 a_1^\dagger a_1 + E_2 a_2^\dagger a_2  + J (a_2^\dagger a_1 + a_1^\dagger a_2).
\end{align}
The terms $E_1$ and $E_2$ are the transition energies of sites 1 and 2, and $J$ is the pigment-pigment coupling energy.
We restrict the system to at most one excitation on each site, so our state space is spanned by the set $\{\ket g, \ket{e_1}, \ket{e_2}, \ket f\}$, denoting the ground state, excitation of the first or second site, and excitation of both sites.
We further assume the usual linear coupling to a bath of phonons.
Details of the bath are specified below.
The electronic part of this Hamiltonian can be diagonalized by applying a unitary rotation $U$ to the single-excitation subspace, given by
\begin{align}
	U = \begin{bmatrix} \cos \theta & \sin \theta \\ -\sin\theta & \cos\theta \end{bmatrix},
\end{align}
where we defined the mixing angle $\theta = \frac{1}{2}\arctan(2J/\Delta)$ with $\Delta = E_1 - E_2$. These single excitation eigenstates are denoted $\ket\alpha$ and $\ket\beta$,
The transition dipole moments for each pigment are $\vec d_1$ and $\vec d_2$, oriented with relative angle $\phi$.

\subsection{Analytical calculation of pump-probe response}

To begin, we choose a parametrization for the excited state density matrix of a dimer. In general, any valid density matrix for a two-level system can be written in any basis in form $\frac{1}{2}(I +  \vec r \cdot \vec\sigma)$, in terms of the Pauli matrices $\vec \sigma = \{\sigma_x, \sigma_y, \sigma_z\}$ and the Bloch vector $\vec r = \{r_1,r_2,r_3\}$, with $r_i$ real and $|\vec r|^2 \leq 1$ \cite{Nielsen2000}. This can be straightforwardly generalized to unnormalized density matrices by adding the normalization $r_0$ and defining $\sigma_0 = I$, in which case the set of valid but unnormalized states are those that can be written in the form
$\frac{1}{2}  (\vec r \cdot \vec \sigma)$, where $\vec r$ is now the 4-dimensional vector $\{r_0, r_1,r_2,r_3\}$ with constraints $r_1^2 + r_2^2 + r_3^2 \leq r_0^2$ and $r_0 > 0$.
We will use these four real parameters to parametrize the excited state electronic density matrix $\rho_e$ for our tomography protocol, since it has population $r_0 \ll 1$.
Using this representation, the total second-order correction to the electronic density matrix from Eq.~\eqref{eq:rhopp-rhoe} is given by
\begin{align}
	\rhopp = -r_0 \ketbra{g}{g} + \frac{\vec r \cdot \vec \sigma}{2}.
	\label{eq:rho2_r}
\end{align}
For convenience, we suppose the state is written in terms of the eigenbasis expansion of the excited states $\{\ket\alpha, \ket\beta\}$, so the parameters $r_1$ and $r_2$ correspond to excitonic coherences and $r_3$ corresponds to the balance of population between excitonic states.
In practice, the experimental signal is only known up to a constant factor, so we can only hope to be able to reliably determine the normalized excited-state density matrix, given by the usual Bloch-vector elements $\{r_1/r_0, r_2/r_0, r_3/r_0\}$.

It is now straightforward (if tedious) to write down the exact pump-probe response function in terms of microscopic parameters.
For illustrative purposes, we do so here for a dimer with a Markovian bath described by Redfield theory in the secular approximation \cite{May2011}. The time-evolution contained directly in the pump-probe response function is for times following the probe interaction, so the relevant part of the system density matrix for this evolution includes coherences between ground and singly excited states and between singly and doubly excited states.
In secular-Redfield theory, coherences in the excitonic basis only evolve with exponential decay, $G(T) \ketbra{a}{b} = e^{-\gamma_{ab} T} \Theta(T) \ketbra{a}{b}$,
 where $G(T)$ is the retarded material Green function denoting evolution for time $T$, $\Theta$ is the Heaviside step function, and $\gamma_{ab}$ is some complex number with positive real part. Since the formulas for the response function will accordingly be most compact in the exciton basis, we consider the excitonic transition dipole moments given by,
\begin{subequations}
\begin{align}
	\mu_{g\alpha} &= \mu_1 \cos\theta + \mu_2 \sin\theta \\
	\mu_{g\beta} &=  -\mu_1 \sin\theta + \mu_2 \cos\theta \\
	\mu_{\alpha f} &= \mu_1 \sin\theta + \mu_2 \cos\theta \\
	\mu_{\beta f} &=  \mu_1 \cos\theta - \mu_2 \sin\theta,
\end{align}
\end{subequations}
with $\mu_i = d_i a_i$, where $d_i$ is the component of the dipole-transition vector parallel to the probe polarization.
For convenience, we define $f_\alpha$ and $f'_\alpha$ to denote the Fourier transform of the time evolution operator that leads to a peak in the pump-probe spectrum at frequency $\omega_\alpha$, with decay constant $\gamma$ or $\gamma'$, where the prime indicates the decay constant for the transition between the 1- and 2-exciton manifolds instead of between the 0- and 1-exciton manifolds:
\begin{align}
	f_\alpha = \frac{1}{i (\omega_\alpha - \omega) - \gamma},
	\qquad f'_\alpha = \frac{1}{i (\omega_\alpha - \omega) - \gamma'}.
\end{align}
We define $f_\beta$ and $f'_\beta$ analogously, for the components peaked at $\omega_\beta$.

Using Eq.~\eqref{eq:pp-response-vec}, the calculation of the pump-probe response function for an arbitrary state is determined by the vectorized version of the pump-probe operator,
$\Lbra{\mathcal{P}(\omega)}$. For this dimer problem, we define the pump-probe bra vector such that $\Rpp = \Lbraket{\mathcal{P}}{r}$, where $\Lket{r} = \vec{r} = \{r_0, r_1, r_2, r_3\}$. Such a relation still holds upon substitution of $\Lket{r}$ for $\rhopp$, since the relation between the two given by Eq.~\eqref{eq:rho2_r} is linear.
With this convention, evaluating the pump-probe operator in Eq.~\eqref{eq:pp-response-operator}
for this dimer model described by secular-Redfield theory yields the general result,
\begin{align}
	\Lbra{\mathcal{P}}
	\propto
	\begin{bmatrix}
		-3 \muga^2 f_\alpha + \mubf^2 f_\alpha^\prime
		-3 \mugb^2 f_\beta + \muaf^2 f_\beta^\prime \\
		-\muga\mugb \left(f_\alpha + f_\beta\right) + \muaf\mubf \left(f_\alpha^\prime + f_\beta^\prime\right) \\
		-i \left[\muga\mugb \left(f_\alpha - f_\beta\right) + \muaf\mubf \left(f_\alpha^\prime - f_\beta^\prime\right)\right] \\
		-\muga^2 f_\alpha - \mubf^2 f_\alpha^\prime
		+ \mugb^2 f_\beta + \muaf^2 f_\beta^\prime
	\end{bmatrix}^T.
	\label{eq:dimer-projector}
\end{align}
This equation holds for each single molecule that would contribute to the pump-probe signal.
We can also calculate the exact isotropic average of Eq.~\eqref{eq:dimer-projector} over an ensemble of randomly oriented molecules.
In terms of the original Hamiltonian parameters, it is given by
\begin{widetext}
\begin{align}
	\stretchiso{\Lbra{\mathcal{P}}}
	\propto
	\begin{bmatrix}
	 \left(\cos ^2 \theta + \delta^2 \sin ^2 \theta \right) (f'_{\alpha }-3 f_{\alpha })+ \left(\sin ^2\theta + \delta^2 \cos ^2\theta \right) (f'_{\beta }-3 f_{\beta }) + \delta  \sin 2 \theta \cos \phi \left(-f'_{\alpha
	   }+f'_{\beta }-3 f_{\alpha }+3 f_{\beta }\right) \\
	 - \frac{1}{2} \left(\delta^2-1\right) \sin 2\theta \left(f'_{\alpha }+f'_{\beta }+f_{\alpha}+f_{\beta }\right) + \delta  \cos 2\theta \cos\phi \left(f'_{\alpha }+f'_{\beta }-f_{\alpha }-f_{\beta }\right) \\
	 i \left[\frac{1}{2} \left(\delta^2-1\right) \sin 2 \theta \left(f'_{\alpha }-f'_{\beta }+f_{\alpha }-f_{\beta }\right)+ \delta  \cos 2\theta \cos \phi  \left(-f'_{\alpha }+f'_{\beta
	   }+f_{\alpha }-f_{\beta }\right)\right] \\
	 -\left(\cos ^2 \theta + \delta^2 \sin ^2 \theta \right) (f'_{\alpha }+ f_{\alpha })+ \left(\sin ^2\theta + \delta^2 \cos ^2\theta \right) (f'_{\beta } + f_{\beta }) + \delta \sin 2 \theta \cos \phi \left(f'_{\alpha }+f'_{\beta
	   }-f_{\alpha }-f_{\beta }\right) \\
	\end{bmatrix}^T,
	\label{eq:dimer-projector-iso}
\end{align}
\end{widetext}
where $\theta$ is the excitonic mixing angle, $\delta = |\vec d_2|/|\vec d_1|$ is the ratio of the two site transition dipole moments and $\phi$ is the angle between them. Note that neither of these equations includes the effects of static disorder, which could be accounted for by averaging the pump-probe response function over each member of the ensemble. Formally, it does not suffice to separately average $\Lbra{P}$, since under static disorder the state $\Lket{r}$ also varies in correlated way (see Sec.~\ref{sub:response_function}).

Equation~\eqref{eq:dimer-projector-iso} makes it possible to determine some cases in which inverting the isotropically averaged state cannot possibly be successful, regardless of the exact inversion protocol. We can identify these cases because successful inversion requires that the elements of $\iso{\left\langle\!\middle\langle \mathcal{P} \right|}$ be linearly independent.
For example, in the homodimer case with both pigments fixed to have the same transition energies ($\theta = \pi/4$ or $\theta = 3\pi/4$) and equal transition-dipole moment magnitudes ($a = 1$), the pump-probe signal does not depend on the coherence terms, so it will be impossible to invert them ($r_1$ and $r_2$). Likewise, the coherence terms do not contribute if the transition dipole moments have identical magnitude ($\delta = 1$), and either are oriented perpendicularly ($\cos\phi= 0$) or there are matching dephasing rates for the 0-1 and 1-2 coherences ($f_\alpha = f'_\alpha$ and $f_\beta = f'_\beta$, as occurs in the high-temperature limit).
As Yuen-Zhou \emph{et al.}\ found for the same dimer model~\cite{Yuen-Zhou2011}, quantum process tomography also fails under similar but not identical conditions.

\subsection{Numerical example}

For a numerical example, we consider the dimer model used in a prior investigation of quantum process tomography~\cite{Yuen-Zhou2011}. We model excitation by a resonant \SI{40}{fs} full-width-at-half-maximum (FWHM) pump centered at \SI{12800}{cm^{-1}}.
The parameters in the electronic Hamiltonian are $E_1 = \SI{12881}{cm^{-1}}$, $E_2 = \SI{12719}{cm^{-1}}$ and $J = \SI{120}{cm^{-1}}$, and the experiment is performed on an ensemble with normally distributed static disorder of standard deviation \SI{40}{cm^{-1}} added to each site energy. The transition dipole moments are fixed with ratio $\delta = |\vec{d}_2/\vec{d}_1| = 2$ and orientation angle $\phi=0.3$. Each pigment is assumed to be coupled to an independent bath of phonons, with spectral density of the form $J(\omega) = \frac{\lambda}{\omega_c}\omega e^{-\omega/\omega_c}$ with $\omega_c = \SI{120}{cm^{-1}}$ and $\lambda = \SI{30}{cm^{-1}}$. The bath is assumed to be at thermal equilibrium at $T = \SI{273}{K}$ and is modeled by secular Redfield theory \cite{May2011}, including only the real (dissipative) part of the Redfield tensor.

\begin{figure}[]
	\includegraphics[]{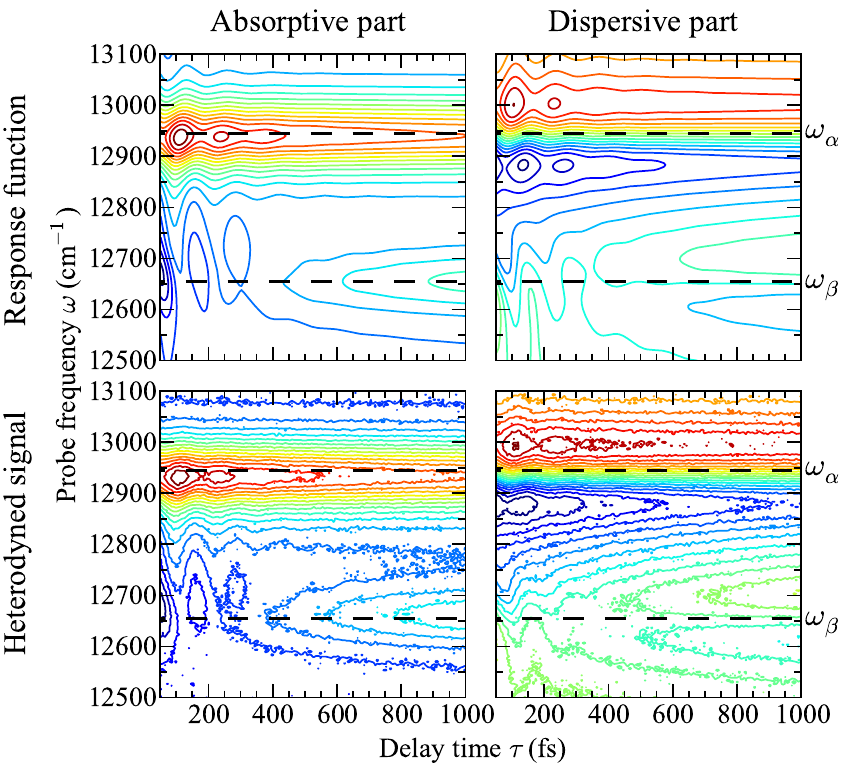}
	\caption{\label{fig:pp-response}Absorptive (left) and dispersive (right) parts of the pump-probe response function $\Rpp(\omega, \rhopp(\tau))$ (top) and the corresponding heterodyne detected signal $S(\omega,\tau)$ (bottom) for our dimer model system. The dashed line indicates the two exciton transition energies in this system. Only the absorptive part (left) is revealed directly by a pump-probe experiment. Obtaining the dispersive part (right) requires a transient grating setup with heterodyne detection, as described in Sec.~\ref{sub:detection_scheme}.}
\end{figure}

To simulate an experimental dataset, we first calculate the non-linear polarization $P^{(3)}(\omega,T)$ for a probe of the same shape as the pump pulse, on a grid of 181 probe frequencies $\omega$ (intervals of \SI{3.33}{cm^{-1}} between \SI{12500}{cm^{-1}} and \SI{13100}{cm^{-1}}) and 140 central time-delays $T$ between pump and probe pulses (intervals of \SI{6.81}{fs} between \SI{50}{fs} and $\SI{1}{ps}$).
From this non-linear polarization, we then calculate the results of a hypothetical heterodyne detection with a local oscillator matching the probe pulse, with and without a $\pi/2$ phase shift. Finally, we accounted for noise in detection by including additive noise with uniformly random phase and amplitude drawn from a standard deviation with width equal to $10^{-2}$ times the maximum amplitude over all delay times and frequencies of the heterodyne detected signal $S(\omega,T)$.
These simulated measurements, generated for comparison both with and without detection noise, are shown in Fig~\ref{fig:pp-response}, together the response function from which they are calculated.

\subsection{Response function inversion}
\label{sec:response-inversion-dimer}

\begin{figure}[]
	\includegraphics[]{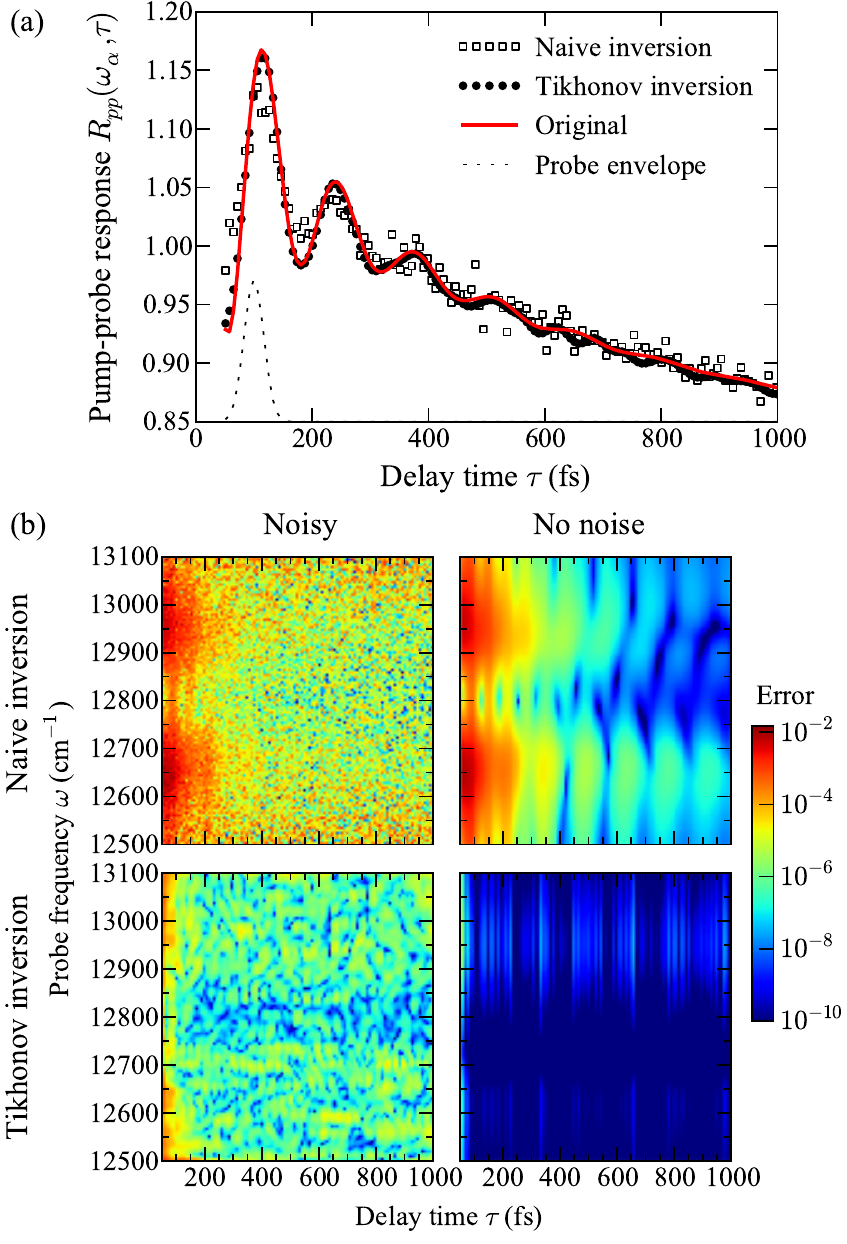}
	\caption{\label{fig:deconvolution-example}(a) Example reconstruction of the pump-probe response at fixed probe-frequency $\omega_\alpha$ for an instance of the high-noise test problem. (b) Errors in the inverted pump-probe response obtained by the direct and Tikhonov inversion methods for a single example of the low and high noise test problems. The error is given by the absolute value squared of the difference between the estimated and actual response function, $|\hat{R}_\text{pp}(\omega,\tau) - \Rpp(\omega,\tau)|^2$.}
\end{figure}

Figure~\ref{fig:deconvolution-example} illustrates the performance of the completed double Tikhonov regularization based deconvolution algorithm for typical noisy and noise free examples of our test problem.
We compare with the ``naive'' approach of assuming that the probe is impulsive and using Eq.~\eqref{eq:naive-inversion} to obtain the response function by simply dividing the signal by absolute value squared of the probe field $|E_\text{pr}(\omega)|^2$.
Table~\ref{tab:deconvolution-summary} summarizes the results of the simulated inversion for the noise free case and 1000 such noisy examples.
\begin{table}[]
	\begin{ruledtabular}
	\begin{tabular}{ccccc}
	Noise & $S \to P^{(3)}$ & $P^{(3)} \to \Rpp$ & RMSE & Improvement \\ \hline\hline
	$10^{-2}$ & Naive & Naive & $12.12 \pm 0.07$ & --- \\
	$10^{-2}$ & Tikhonov & Naive & $7.78 \pm 0.05$ & $1.6 \pm 0.0$ \\
	$10^{-2}$ & Naive & Tikhonov & $3.34 \pm 0.08$ & $3.6 \pm 0.1$ \\
	$10^{-2}$ & Tikhonov & Tikhonov & $0.98 \pm 0.06$ & $12.5 \pm 0.7$ \\
	\hline
	0 & Naive & Naive & 7.110 & --- \\
	0 & Tikhonov & Naive & 7.110 & 1.0 \\
	0 & Naive & Tikhonov & 0.005 & 1301 \\
	0 & Tikhonov & Tikhonov & 0.005 & 1301 \\
	\end{tabular}
	\end{ruledtabular}
	\caption{\label{tab:deconvolution-summary}Summary of deconvolution performance over 1000 instances of simulated experimental noise. RMSE (root-mean-squared-error) is given by the sum of the absolute difference between the estimated and actual response functions, $\left(\sum_{\omega,\tau} |\hat{R}_\text{pp}(\omega,\tau) - \Rpp(\omega,\tau)|^2\right)^{1/2}$. Improvement is the multiple of the reduction in RMSE compared to the naive approach. Uncertainties indicate one standard deviation in the empirical distribution.}
\end{table}
In addition to the double Tikhonov and naive methods, we also consider the alternatives of substituting the naive approach individually for each of the two Tikhonov steps. Recall that in the first stage of the inversion ($S \to P^{(3)}$), the Tikhonov regularization serves only to smooth the data. It is not surprising then that Table~\ref{tab:deconvolution-summary} shows that the specific method chosen for this first stage (i.e., naive or Tikhonov) makes no difference for the noise free case. In the second stage ($P^{(3)} \to \Rpp$), the Tikhonov regularization also performs a deconvolution over the probe envelope.

The results in Fig.~\ref{fig:deconvolution-example} and Table~\ref{tab:deconvolution-summary} show that the Tikhonov based inversion is a clear improvement over the naive approach, reducing the root-mean-squared-error (RMSE) by a factor of 12 for our noisy example and 1300 for our noise-free example.
The noisy and noise free examples allow us to observe that the Tikhonov regularizations remove two types of errors inherent in the naive inversion: (1) errors from noisy measurements and (2) errors associated with the convolution of the pump-probe response over the finite probe duration. In the noisy case, both errors are large; in the noise free case, there are only errors from the second source. Clearly, reducing the experimental noise associated with measurement alone does not suffice to accurately estimate the pump probe response: as the double Tikhonov inversion of the noisy signal outperforms naive inversion of the noise free signal by a factor of 7 in RMSE.

As Figure \ref{fig:deconvolution-example} shows, the errors in the estimate of the response function are not uniformly distributed, revealing structure relevant to our specific example and also to more generic systems.
The largest errors are associated with smallest delay times. This makes sense, since the smallest delay times are those at which at the response function (shown in Fig.~\ref{fig:pp-response}) varies most rapidly.
For almost any system, the pump-probe spectrum will change fastest at short delay times, but this is especially true for our example system, where the pump-probe signal includes contributions from quickly oscillating coherences.
The Tikhonov estimates face an additional stability challenge at short delays times, since, as discussed above, the reconstruction cannot use measurements from the pulse overlap regime.

\subsection{State tomography}

Since we have demonstrated that the first, response function inversion can be performed with vanishing error, we now consider inverting the exact pump-probe response function to obtain the state of our model dimer. Despite the presence of static disorder in our example, we use the factorization of the response function in Eq.~\eqref{eq:pp-response-measure-iso}. We are obliged to
do so even though strictly speaking the relationship does not hold, because the alternative of reconstructing the density matrix for each member of the ensemble from a bulk measurement is unrealistic. Accordingly, even without adding noise associated with the measurement, when carried out for an ensemble, our inversion faces potential stability issues because of the static disorder.

The degeneracies and near degeneracies in Eq.~\eqref{eq:dimer-projector-iso} mean that for most Hamiltonian choices our inversion algorithm can only robustly extract at most three of the four parameters necessary to fully characterize the dimer excited state, since the reconstruction matrix will be poorly conditioned. The condition number of a linear transformation gives a bound on the multiplicative increase in the relative error after performing the linear transformation \cite{Hansen:2002ut}. For our specific numerical example, the condition number drops from 3700 to 3.1 when we include only three parameters.
One source of these stability issues for a dimer is evident from Eq.~\eqref{eq:dimer-projector-iso}: since our numerical example has well separated transition energies, the main contribution to the peaks in the dispersive part of the signal is to the imaginary part of the coherence term, $r_2$. This leaves our inversion to recover three parameters ($r_0$, $r_1$ and $r_3$) from the two peak amplitudes in the absorptive signal. Since recovering three unknowns from two equations is not possible, we need to fix one of these values in order to make the inversion stable. The obvious choice is to fix the normalization $r_0$, since the total excited state population should remain constant after the end of the pump pulse until spontaneous decay, on timescales approaching \SI{1}{ns} for natural pigment-protein complexes \cite{Blankenship2002}. To determine the normalization, we solve for it at a moderately long delay time (e.g., $\tau = \SI{10}{ps}$) at which point we can safely assume (at least under secular Redfield dynamics) that the real part of the coherence $r_1 \to 0$, but very few excitations have been lost. If these timescales are not easily separable, then this normalization term could be fit to an exponential decay.

\begin{figure}[t]
	\includegraphics[]{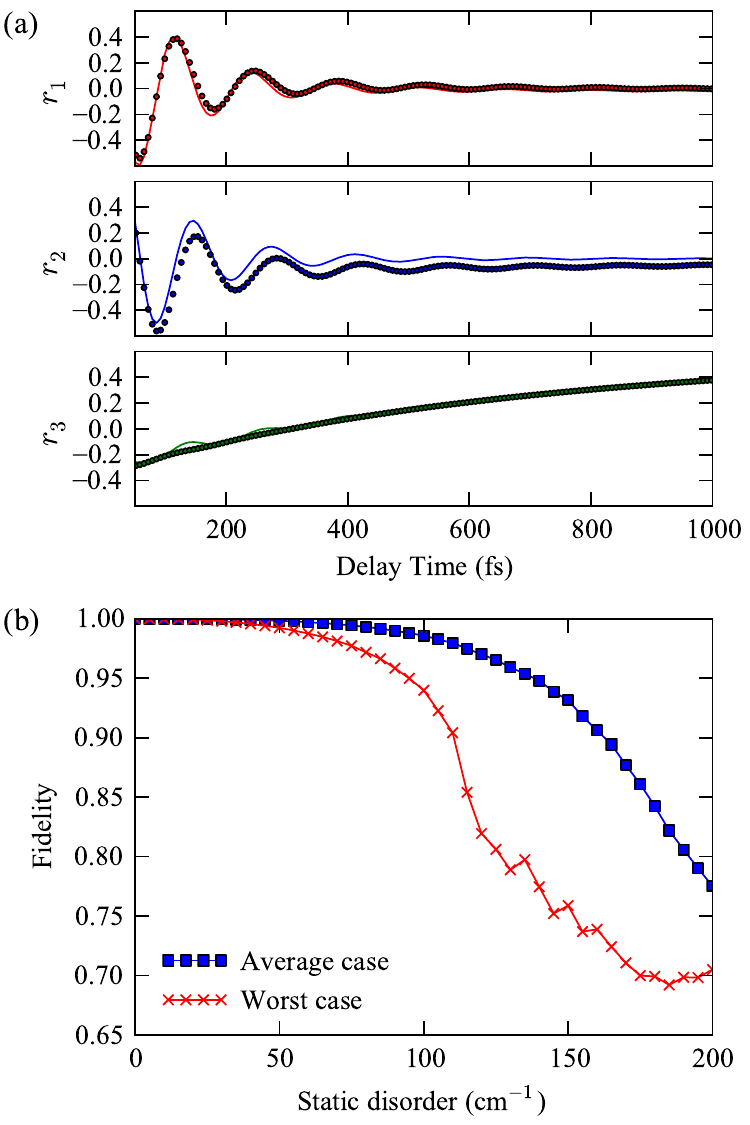}
	\caption{\label{fig:qst-example}Results of quantum state tomography for our dimer test problem. (a) Original (solid) and reconstructed (dotted) values for each element of the Bloch state vector for the reconstruction with static disorder of standard deviation \SI{40}{cm^{-1}}. Normalization is omitted since the state vector elements are rescaled such that $r_0 = 1$ fixed for all times following initial excitation. (b) Worst- and average-case fidelities for the reconstructions $\rho_e(\tau)$ for delay times $\tau$ in the range \SI{50}{fs} to \SI{1}{ps} as a function of the width (standard deviation) of the distribution of static disorder. Results are obtained from an ensemble average over $10^6$ samples for each point.}
\end{figure}

The results of applying this state tomography procedure to our numerical example with varying levels of static disorder are shown in Fig.~\ref{fig:qst-example}.
The fidelity, ranging from 0 to 1, provides a numerical summary of the quality of the state reconstruction \cite{Nielsen2000}.
For the level of static disorder chosen by Yuen-Zhou \emph{et al.}\ (\SI{40}{cm^{-1}}), the reconstruction [panel (a)] has a worst-case fidelity of $99.5\%$ over delay times $T$ shorter than \SI{1}{ps}, and an average-case fidelity of $99.9\%$. However, we see that both the worst-case and average-case fidelities drop sharply as the static disorder is increased above this level [panel (b)], since our assumption that the pump-probe response can be factorized between the pump-probe projection and the second order density matrix becomes increasingly unrealistic.

\section{Scaling to larger systems}
\label{sec:scaling_to_larger_systems}

Can state tomography scale to systems larger than an excitonic dimer? In particular, can we apply it to precisely reveal the excitonic state in a natural light-harvesting system? Any scaling difficulties will be encountered in the second step of our inversion protocol, to invert the quantum state from the response function, since the relationship between the response function and measured signal does not directly depend on system parameters.
Successful state tomography certainly requires both knowledge of how each density matrix element contributes to measurements and appropriate conditions such that each element, at least in principle, makes an independent and non-zero contribution.
By construction, these conditions were satisfied for our hypothetical dimer example.
We found that the primary limitation on inverting the state was ensemble disorder, which can in principle be avoided by single molecule techniques.
Now, to explore the limits of state tomography, we relax these assumptions in order to consider the feasibility of state tomography in an actual protein-pigment complex.

As a model light-harvesting system, we focus on a monomer of the Fenna-Matthews-Olson (FMO) complex of green sulfur bacteria, which consists of 7 pigment molecules \cite{Adolphs2006, Savikhin1997}. The FMO complex is a widely used model system for understanding photosynthetic energy transfer and is thus is one of the best characterized natural protein-pigment complex. The crystal structure for the FMO complex is known, which combined with input from spectroscopy experiments, has allowed for general agreement on an electronic Hamiltonian \cite{Adolphs2006, Hayes2011}. Because the arrangement of pigments is fairly disordered, each excited state in a monomer of the FMO complex is bright, although they overlap in the presence of homogeneous and inhomogeneous broadening. This is important, since full state tomography would certainly not be possible on a system with multiple dark states, because no optical probes could reveal the distribution of energy among those states. However, typical of the situation for other natural pigment-protein systems, there is little consensus on the magnitude of the static disorder or the spectral density of the electronic-vibrational coupling \cite{Shim:2012el}. These difficulties are compounded by the theoretical and computational challenge of modeling dynamics in a system as large as FMO exactly for arbitrary system-bath interaction strength \cite{Ishizaki2009a}. For our concrete example, we use the electronic Hamiltonian for FMO of \emph{Chlorobaculum tepidum} from Ref.~\citenum{Adolphs2006}, with the spectral density and computational model of secular Redfield theory matching those used in for the dimer example. This model includes only one electronic state per pigment and assumes Gaussian distributed static disorder with standard deviation \SI{42.5}{cm^{-1}} (\SI{100}{cm^{-1}} FWHM).

Our formalism for the pump-probe response function allows us to place bounds on the feasibility of any state tomography procedure, since the relationship between system information and the resulting pump-probe spectra is entirely contained in the pump-probe response operator $\mathcal{P}(\omega)$.
By looking at the ensemble average of this operator, we implicitly consider inversion under the scenario that the average over static disorder can be factorized between the pump-probe operator and the density matrix,. This assumption was successful in the dimer example above when the magnitude of static disorder was not too large.
As discussed in Section~\ref{sub:response_function}, the pump-probe operator at each frequency can be interpreted as a Liouville space bra-vector $\Lbra{\mathcal{P}(\omega)}$. Accordingly, it is possible to interpret the calculation of a pump-probe response as the act of applying the linear operator
\begin{align}
	\mathbb{P} = \int d\omega \ket{\omega} \Lbra{\mathcal{P}(\omega)}
	\label{eq:Poperator}
\end{align}
to the state $\Lket{\rhopp}$. We now consider the properties of the linear operator $\mathbb{P}$ in the limit of effectively continuous sampling of probe frequencies $\omega$. To represent states in Liouville space, we use a basis set that allows us to represent each state with $n^2$ real values, in terms of populations $\ketbra{n}{n}$ and coherences $\ketbra{n}{m} + \ketbra{m}{n}$ and $i\ketbra{n}{m} - i\ketbra{m}{n}$.
This allows us to construct a real-valued version of the map $\mathbb{P}$ that
takes real valued state vectors to real valued spectra by concatenating the real and imaginary parts of $\mathbb{P}$.

\begin{figure*}[t]
	\includegraphics[]{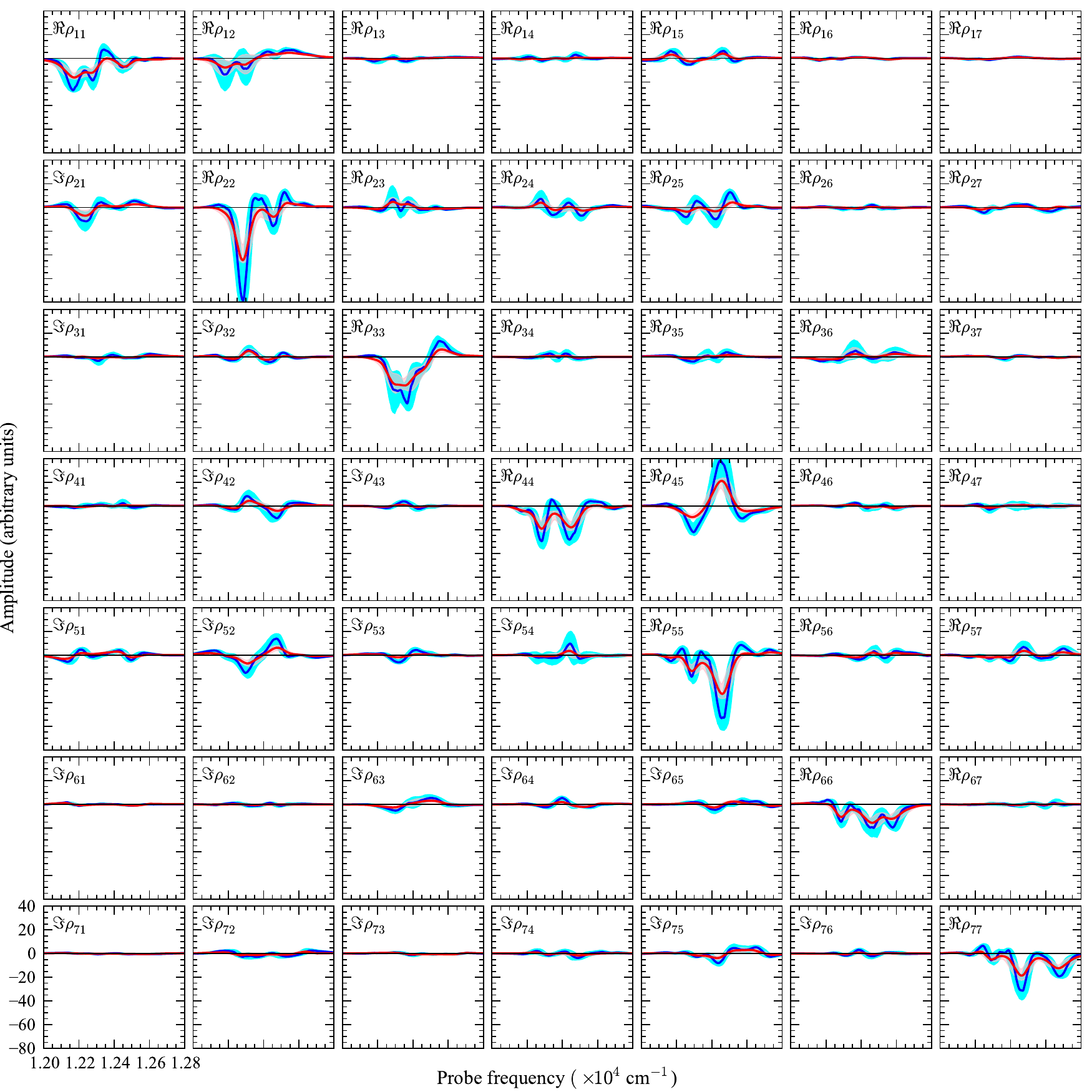}
	\caption{\label{fig:species_assoc_spectra}Species associated spectra, defined by the contribution of the marked density matrix elements to the pump-probe response, for the FMO complex at \SI{77}{K} (blue) and \SI{300}{K} (red), obtained as the average of 1000 samplings over static disorder. Labels indicate the contributing density matrix element in the excitonic basis. Shaded regions indicate central 95\% confidence intervals obtained from 1000 additional samplings over Hamiltonian uncertainty, as described in the text.}
\end{figure*}

To begin, in Figure~\ref{fig:species_assoc_spectra} we plot the elements of the absorptive (real) part of the pump-probe operator for our model of the FMO complex in the isotropic average (magic angle configuration).
We represent the operator in terms of the species associated spectra of each exciton population and the real and imaginary part of each coherence, so that the pump-probe spectra of any excited state is equal to the linear combination of the plotted spectra weighted by the indicated density matrix elements. In addition to the unperturbed spectra, we also plot the range of possible spectra given current uncertainty about the best fit parameters. We conservatively estimate the uncertainty in the electronic Hamiltonian by sampling over additive independent Gaussian noise of width \SI{20}{cm^{-1}} for each site energy and 10\% of the value of each off-diagonal coupling. This uncertainty is in addition to the static disorder, which we leave with fixed magnitude. The most striking feature of these spectra is that, at least in the isotropic average, the dominant contribution to the pump-probe spectra is from the population terms. The smaller contribution of most coherence terms, compounded by the already smaller values of the coherences in the density matrix due to dephasing, explains why it is difficult to observe oscillations due to electronic coherence in pump-probe spectra \cite{Savikhin1997}. Even for extremely precise measurements, the uncertainty in some of these species associated spectra suggests that our current Hamiltonian characterization does not suffice to reliably obtain most density matrix elements. Indeed, the dominance of the diagonal terms suggests that a practical scheme for partial state tomography could consist of entirely ignoring the off-diagonal terms.

\begin{figure}[]
	\includegraphics[]{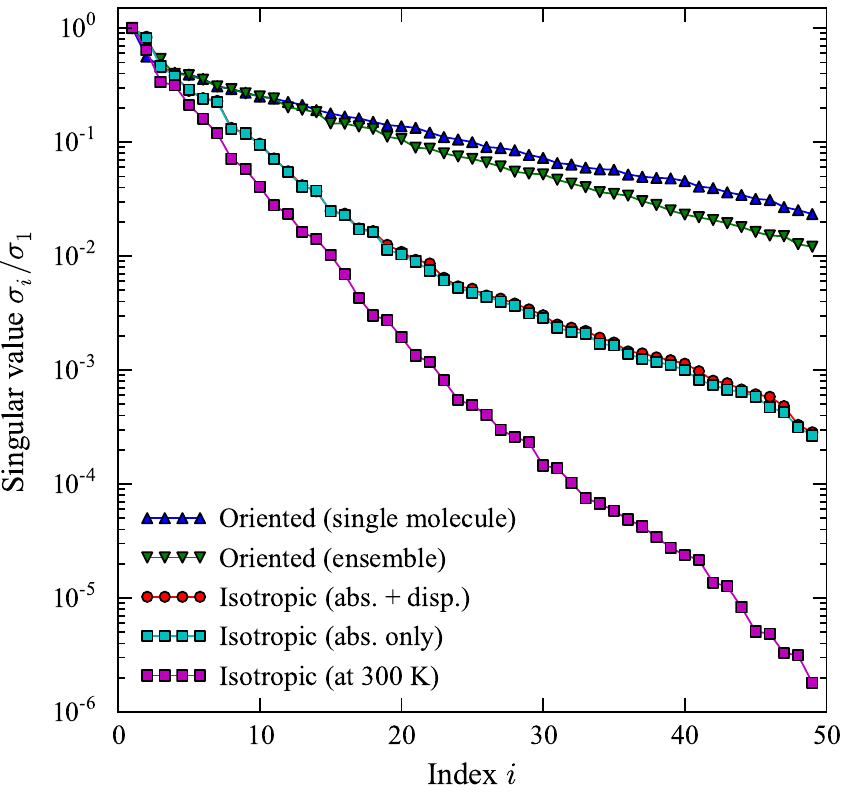}
	\caption{\label{fig:singular-values}Normalized singular values from a singular value decomposition of the real valued pump-probe map $\mathbb{P}$ given by Eq.~\eqref{eq:Poperator} for the FMO complex under various conditions. The top line is from the spectra of a single monomer at \SI{77}{K}, from the combination of measurements in all independent polarization configuration. Subsequent lines add additional constraints, which apply cumulatively: ensemble measurement (over static disorder), the isotropic average of the signal, only the absorptive (real) part of the signal and finally performing the measurement at room temperature.}
\end{figure}

Another approach to estimating the feasibility of inversion for arbitrary states is to look at the spectral properties of the operator $\mathbb{P}$ as revealed by the singular value decomposition, $\mathbb{P} = U S V^\dagger$, where $U$ and $V$ are unitary and $S$ is diagonal with positive elements. In particular, we focus on the singular values $\sigma_i$, given by the diagonal elements of $S$ in descending order and normalized to the highest singular value $\sigma_1$.
To compare the feasibility of inversion under various conditions, we plot these singular values in Figure~\ref{fig:singular-values}.
The singular values reveal significant information about the feasibility of an inversion: in general, inversion is more feasible when the singular values $\sigma_i$ decay more slowly \cite{Hansen:2010vw}. For example, the condition number, which gives an upper bound on the ratio by which the relative error can increase in an inversion, is equal to the ratio of the largest to the smallest singular values. In Figure~\ref{fig:singular-values}, the condition number is one over the value shown for $i=49$.

\begin{figure}[]
	\includegraphics[]{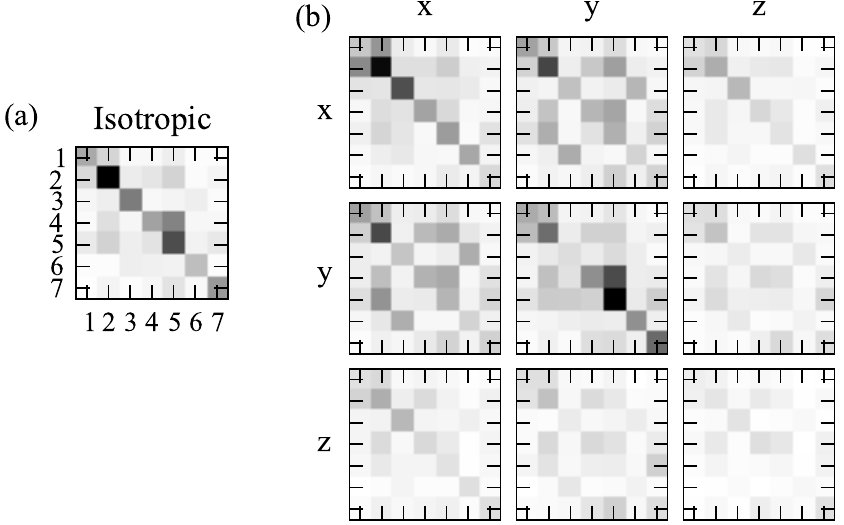}
	\caption{\label{fig:species-max-amp}Maximum amplitude over probe frequencies of the species associated spectra for an FMO monomer at \SI{77}{K} for (a) the isotropic average and (b) each independent polarization configuration of the probe and local oscillator, including the ensemble average over static disorder. The labeling of each species matches that used in Figure~\ref{fig:species_assoc_spectra}: all entries including and above the diagonal correspond to the real part of the matching density matrix element (in the excitonic basis), and all entries below the diagonal correspond to the imaginary part. The cartesian coordinates were chosen arbitrarily, matching those used in an assignment of the crystal structure.}
\end{figure}

Figure~\ref{fig:singular-values} provides an indication of the relative significance of different experimental constraints insofar as they affect state tomography.
We see two major changes that reduce the condition number of the inversion by around two orders of magnitude each. First, reducing temperature from \SI{300}{K}  to \SI{77}{K} improves the conditioning because spectral features in the species associated spectra are sharpened. Second, changing from a randomly oriented (isotropic) to an oriented sample helps because in principle 9 times more independent measurements are possible than in the magic angle configuration, one for each combination of $x$, $y$ and $z$ polarizations for the probe and signal interactions. This is illustrated in Figure~\ref{fig:species-max-amp}, which plots the peak amplitudes of each of the species associated spectra. In some cross-polarization configurations, coherences give contributions of similar magnitude to populations.
Single molecule measurements do offer an important advantage over oriented ensembles that is not seen in this figure, since their analysis does not require any assumption that the average over static disorder can be factorized between pump and probe.
Finally, this figure suggests that the dispersive component of the signal, which as discussed in Section~\ref{sub:detection_scheme}
requires a transient grating experiment, offers little additional information
compared to the absorptive component that is provided directly by the pump-probe signal.

\section{Conclusions}

What information does an ultrafast spectroscopy experiment tell us about an excitonic system? How can we best design these experiments?
We believe that a powerful way to answer these questions is to treat ultrafast spectroscopy as a hierarchy of statistical inverse problems, as we have demonstrated here for inverting the electronic excited state.
By dividing our analysis into many smaller steps---from experimental signal to response function to excited state and eventually on to dynamics, equations of motion and Hamiltonian parameters---we can see exactly how and where our models and experiments fall short.
In this regard, our approach contrasts strongly with the established procedure of ``forward simulation'' for determining Hamiltonian parameters in complex excitonic systems by simultaneously fitting many experiments with an assumed theoretical model for calculating spectra from first principles \cite{Adolphs2006,Novoderezhkin2011}.
Moreover, to identify the time-evolving excited state or another intermediate quantity in our approach, we do not need to introduce additional errors by recalculating from first principles.

Our multi-stage inversion has clear extensions to more general non-linear spectroscopies beyond QST and pump-probe.
As we describe in Appendix~\ref{sec:alternative_formulation}, we could apply essentially the same inversion procedure to a photon-echo experiment to determine the phase-matched component of the 2nd-order density matrix that contributes to the observed signal [Eq.~\eqref{eq:rho_2_PE}]. If it is possible to construct a full set of independent phase-matched initial conditions, our state tomography procedure could be used for process tomography, along the lines of a previous proposal \cite{Yuen-Zhou2011}. More generally, this suggests a new indirect approach for process tomography: first, field information should be used to invert the 3rd-order response function; second, system information should be used to extract the process matrix.
Inverting the response function as an intermediate quantity allows us to be sure we have obtained the maximum information from experiments before considering the harder theoretical problem of extracting system parameters, such as the state, process matrix or underlying Hamiltonian. Such extensions will be pursued in future work.

More immediately, our approach is particularly well suited to evaluating the benefits of employing colored pulses or an ultrafast pulse shaper, because our formalism makes no assumptions about the shape of pump and probe pulses: they are only required not to overlap in time.
In contrast, prior proposals for process tomography relied on the assumption that interactions with laser pulses are much faster than the timescale of dissipative system dynamics \cite{Yuen-Zhou2011, Yuen-Zhou2011a}.
Accordingly, we envision potentially using our scheme for verification of state preparation following shaped pump pulses \cite{Bruggemann2004,Caruso:2012vo}, scenarios for which single molecules or oriented samples are similarly helpful.
Indeed, pump-probe spectroscopy has been used to experimentally verify ultrafast coherent control \cite{Prokhorenko2006}.
The deconvolution step in our inversion protocol also made no assumptions about the shape of the probe pulse. Although our extensive numerical simulations found no cases in which a shaped probe pulse was preferable to the corresponding time-frequency limited pulse, our inversion protocol can just as easily invert the non-linear response from, say, pump-probe measurements where the probe has residual chirp. This suggests the possibility of improving pulse characterization for better time resolution instead of or in addition to efforts to further compress pulses in time \cite{Polli:2010hr}.

Finally, we point out that we have demonstrated state tomography in this work mostly for ensemble systems, including averaging over molecular orientations and static disorder.  These features inevitably reduce the performance of the inversion.
However, both oriented and single molecule experiments may be possible in the near future, given recent advances of preforming ultrafast spectroscopy on crystallized proteins~\cite{Huang:2012vj} and the possiblity of using non-linear florescence measurements with phase-cycling~\cite{Lott2011} to scale non-linear spectroscopy to single molecules. For such single molecule experiments, we expect the present inversion will provide a powerful analytical tool.

\begin{acknowledgments}
We thank Jahan Dawlaty and Yuan-Chung Cheng for helpful discussions.
This work was supported in part by DARPA under award N66001-09-1-2026.
S.H.\ is a DOE Office of Science Graduate Fellow.
\end{acknowledgments}

\appendix

\section{Derivation of pump-probe signal}
\label{sec:response_function}

In third-order spectroscopy, the phase-matched signal depends on the third-order polarization \cite{Abramavicius2009}, which can be written as
\begin{align}
	P^{(3)}_{\vec{k}_s}(t) =& \iiint_0^\infty dt_3\ dt_2\ dt_1\ R^{(3)}_{\vec k_s}(t_3, t_2, t_1) \notag \\
	&\times {E}_3^{u_3} (t - t_3) {E}_2^{u_2} (t - t_3 - t_2) \notag \\
	&\times {E}_1^{u_1} (t - t_3 - t_2 - t_1)
	\label{eq:P3_general}
\end{align}
assuming that the interactions happen in the numbered order and invoking the rotating wave approximation.
The quantity $\vec k_\text{S} = u_1 \vec k_1 + u_2 \vec k_2 + u_3 \vec k_3$
is the signal wave-vector, $(u_1, u_2, u_3) \in \{ (-, +, +), (+, -, +), (+, +, -) \}$ correspond to the three experimental geometries with non-zero signal (rephasing, non-rephasing and double-quantum-coherence, respectively) and ${\rm E}_i^{+}(t)$ denotes the complex profile of the $i$th pulse (we use the convention $E^{-} = (E^{+})^\ast$). The system dynamics are contained in the phase-matched components of the third order response function $R^{(3)}_{\vec k_s}(t_3, t_2, t_1)$, which is given by
\begin{align}
	&R^{(3)}_{\vec k_\text{S}}(t_3, t_2, t_1) = \notag\\ &\left(\frac{i}{\hbar}\right)^3 \tr\left[ \mu^{(-)} G(t_3) V^{u_3} G(t_2) V^{u_2} G(t_1) V^{u_1} \rho_0 \right],
	\label{eq:R3_general}
\end{align}
in terms of the quantities defined in Sec.~\ref{sec:recipe}.

In a pump-probe experiment, the first two interactions are with the same pulse ($\vec k_1 = \vec k_2$), called the pump, and the signal is observed in the direction $\vec k_\text{S} = \vec k_\text{3}$, so $u_1 = -u_2$. The third interaction is with the probe field. Accordingly, the signal is given by adding together the rephasing and non-rephasing interactions, and both pulse orderings 1-2-3 and 2-1-3. Rearranging the terms that result from inserting Eq.~\eqref{eq:R3_general} into Eq.~\eqref{eq:P3_general} yields Eqs.~(\ref{eq:pp-polarization}--\ref{eq:rho2}).

\section{Positivity of $\rho^{(2)}_e$}
\label{sec:properties_of_rho_2}

We consider the system density matrix $\rho$ in the presence of weak fields of strength $\epsilon \ll 1$. Let $\rho^{(n)}$ denote the contribution to the density matrix of strength $O(\epsilon^n)$. Thus we can write $\rho = \rho^{(0)} + \rho^{(1)} + \rho^{(2)} + \rho^{(3)} + O(\epsilon^4)$. Since the temperature is typically several orders of magnitude smaller than the electronic energy gap, the system starts in a tensor product of the electronic ground state $\ketbra{g}{g}$ and the equilibrium vibrational state $\rho_\text{eq}^\text{B}$, $\rho^{(0)} = \ketbra{g}{g} \otimes \rho_\text{eq}^\text{B}$.

We are interested in the excited state portion of $\rhopp$, the component of $\rho^{(2)}$ that contributes to the phase-matched signal $\vec{k}_\text{S} = \vec{k}_\text{pr}$ given by Eq.~\eqref{eq:rho2}. We write this excited state density portion as $\rho^{(2)}_e = \mathcal{Q}\rhopp$, where $\mathcal{Q}$ denotes the projection onto the 1-excitation manifold. Since the non-phase-matched components involve 2-excitation states, we also have $\rho^{(2)}_e = \mathcal{Q}\rho^{(2)}$. This projection is given by $\mathcal{Q}\rho = I_1 \rho I_1^\dagger$, where $I_1 = \sum_m \ketbra{m}{m}$ is the identity operator restricted to the $1$-excitation manifold and $\ket m$ denotes the state where only pigment $m$ is excited. Since the map $\mathcal{Q}$ is written in the appropriate form and $\sum_n I_n I_n^\dagger = I$, $\mathcal{Q}$ is completely positive, with $0 \leq \tr \mathcal{Q}\rho \leq 1$ \cite{Nielsen2000}.

Moving from the ground state to the 1-excitation manifold requires at least two applications of the creation/annihilation operators contained in the dipole transition operators $\mu^{(\pm)}$, and a dipole operator must be applied an even number of times. Accordingly, $\mathcal{Q}\rho^{(n)} = 0$ for $n=0,1,3$, which leaves $\mathcal{Q}\rho = \rho^{(2)}_e + O(\epsilon^4)$. Since  we proved $\mathcal{Q}\rho$ is positive and $\rho^{(2)}$ itself is $O(\epsilon^2)$, we have shown that $\rho^{(2)}_e$ is positive, up to relative errors of $O(\epsilon^2)$.

\section{Tikhonov regularization}
\label{sec:tikhonov_regularization}

The convolution of the pump-probe response with the probe pulse in Eq.~\eqref{eq:convolution} and the non-linear polarization with the local-oscillator in Eq.~\eqref{eq:heterodyne-detect} are
both particular cases of a Fredholm integral equation of the first kind. An extensive literature exists on numerical inversion of such equations, known as discrete inversion problems \cite{Hansen:2010vw}. In general, the discretization of such an integral equation can be written as
\begin{align}
	\vec b = A \vec x + \vec{\epsilon},
	\label{eq:linear-convolution}
\end{align}
where $\vec b$ is the measured signal [e.g., the nonlinear polarization
 $P^{3}(\omega,T)$], $\vec x$ is the desired quantity to invert [e.g., the response function $\Rpp(\omega,T)$]. $A$ is a linear operator representing the integral equation with appropriate coefficients (determined here by the probe field $E_\text{pr}$) and $\vec\epsilon$ denotes some additive experimental error inherent in the data collection.
The obvious solution to estimating $\vec x$ from $A$ and $b$ is to calculate $\vec{\hat x} = A^{-1} \vec b$. However, in practice $A$ may not be invertible.
This is the case for our inversion, since we ignore the experimental signal for times in the pulse overlap regime but still attempt to reconstruct the pump-probe response at those times, which guarantees that $\vec b$ has a lower dimensionality than $\vec x$.
In addition, the presence of even a vanishingly small amount of experimental noise $\vec\epsilon$ makes exact least-squares minimization unsuitable; it will over-fit the noise component $\vec \epsilon$.

Accordingly, to calculate a robust estimate $\hat{\vec{x}}$ of $\vec x$ we use general form Tikhonov regularization \cite{Hansen:2010vw},
\begin{align}
	\hat{\vec{x}} = \argmin_{\vec{x}} \left\{ || A\vec{x} - \vec{b}||^2 + \lambda^2 ||L \vec{x}||^2 \right\}.
	\label{eq:tikhonov}
\end{align}
Tikhonov regularization can be derived formally from the perspective of Bayesian inference, given normally distributed errors and priors \cite{Vogel:2002ta}.
It can be equivalently be expressed as the linear least-squares problem, $\min || [\begin{smallmatrix} A \\ \lambda L \end{smallmatrix}] \vec{x} - [\begin{smallmatrix} \vec{b} \\ 0 \end{smallmatrix}] ||^2 $, and thus the exact solution is given by $\hat{\vec x} = [\begin{smallmatrix} A \\ \lambda L \end{smallmatrix}]^+ [\begin{smallmatrix} \vec{b} \\ 0 \end{smallmatrix}]$, where $+$ denotes the Penrose-Moore pseudoinverse \cite{Hansen:2010vw}.
Ideally, the linear operator $L$ and (real) regularization parameter $\lambda$ are chosen so that $\lambda^2 ||L\vec x||^2$ is an optimally weighted penalty on undesirable features of the solution $\vec x$, reflecting our prior knowledge of the general form of $\vec x$. Common choices of $L$ include the identity matrix $I$ and finite-difference approximations to the first or second derivative given by $(D_1\vec x)_n = x_n - x_{n-1}$ and $(D_2\vec x)_n = x_{n+1} - 2 x_n + x_{n-1}$. We compare different techniques for selecting $\lambda$ and $L$ in Appendix \ref{sec:parameter_selection_techniques}.

There are a number of powerful techniques for calculating efficient approximate solutions to Eq.~\eqref{eq:tikhonov}, especially in cases where the linear operator $A$ is structured \cite{Hansen:2002ut}, such as in our case, where $A$ is a Toeplitz matrix.
For several hundred time-delays or probe frequencies, we find that we can solve
each deconvolution with Eq.~\eqref{eq:tikhonov} exactly and quickly ($\sim$\SI{1}{s} on a modern CPU) by calculating the singular value decomposition of the matrix
$[\begin{smallmatrix} A \\ \lambda L \end{smallmatrix}]$. In principle, it would be possible to solve both steps in the inversion of the pump-probe response in a single two-dimensional Tikhonov regularization. Such 2D inversions are routinely performed in image processing \cite{Hansen:2010vw}, but would require slower, more approximate techniques than the exact solution we used here. Since we find significant improvement without invoking these more complicated methods, we do not use them here.

\section{Parameter selection for Tikhonov regularization}
\label{sec:parameter_selection_techniques}

To perform deconvolutions using Eq.~\eqref{eq:tikhonov}, we need to choose a procedure to select the regularization parameters $\lambda$ and $L$.
In practice, there are a wide variety of techniques for making these selections and the best choice depends on the particular problem at hand \cite{Hansen:2010vw}.
Here we compare the performance of different techniques on simulations matching the dimer problem we analyze in Section \ref{sec:qst_for_a_dimer}.

To begin, we compared the performance of general form Tikhonov regularization with $L$ equal to $I$, $D_1$ and $D_2$, with $\lambda$ chosen optimally so as to minimize the exact mean-squared-error $||\hat{\vec{x}} - \vec{x} ||^2$. A summary of  reconstructions of $\Rpp(\omega, \rhopp(T))$ for 1000 instances of low and high noise is shown in Table~\ref{tab:param-selection}.
\begin{table}[]
	\begin{ruledtabular}
	\begin{tabular}{ccccc}
	Noise & Penalty & Selection & $\lambda_\text{opt}$ & Improvement \\ \hline\hline
	$10^{-2}$ & $I$ & Exact & $0.151 \pm 0.007$ & $2.3 \pm 0.2$ \\
	$10^{-2}$ & $D_1$ & Exact & $0.340 \pm 0.034$ & $4.9 \pm 0.7$ \\
	$10^{-2}$ & $D_2$ & Exact & $0.67 \pm 0.11$ & $6.2 \pm 1.2$ \\
	$10^{-2}$ & $D_2$ & GCV & $0.78 \pm 0.17$ & $5.7 \pm 1.2$ \\
	$10^{-2}$ & $D_2$ & NCP & $2.57 \pm 0.36$ & $2.3 \pm 0.4$ \\
	\hline
	$10^{-3}$ & $I$ & Exact & $0.049 \pm 0.017$ & $5.9 \pm 0.5$ \\
	$10^{-3}$ & $D_1$ & Exact & $0.073 \pm 0.010$ & $30 \pm 5$ \\
	$10^{-3}$ & $D_2$ & Exact & $0.112 \pm 0.017$ & $56 \pm 12$ \\
	$10^{-3}$ & $D_2$ & GCV & $0.103 \pm 0.019$ & $52 \pm 12$ \\
	$10^{-3}$ & $D_2$ & NCP & $0.449 \pm 0.039$ & $19 \pm 2$ \\
	\end{tabular}
	\end{ruledtabular}
	\caption{\label{tab:param-selection}Regularization performance for different penalty operators and parameter selection techniques for 1000 instances of random noise with relative magnitude $10^{-2}$ or $10^{-3}$. Numbers are the mean plus or minus one standard deviation. Improvement is the multiple of the reduction in mean-squared-error for the reconstructed response function using Tikhonov regularization over the error associated with the naive impulse-probe estimate.}
\end{table}
As a benchmark, we consider the ratio of the mean-squared-error from the Tikhonov estimate to the mean-square-error of our naive estimate $\hat R(\omega, \rhopp(T)) \propto P(\omega, T)/E_\text{pr}(\omega)$, which holds in the limit of an instantaneous probe pulse [Eq.~\eqref{eq:naive-inversion}].
For our state inversion algorithm, reconstruction of the response function is most important at frequencies matching the exciton transition energies, so we picked $\omega=\omega_\alpha$, the transition frequency of the higher energy exciton state. We found qualitatively similar results for $\omega=\omega_\beta$ and other choices of $\omega$ as well. As Table~\ref{tab:param-selection} shows, with exact selection of the best regularization parameter $\lambda$, we found consistently best performance with $D_2$, the linear operator approximating the second derivative of the response function with respect to the delay time $T$. This is an intuitively reasonable choice, since plausible response functions should be smooth.

With the choice for $L$ determined, we also need a realistic procedure for selecting the regularization parameter $\lambda$. In a true inversion problem the response function $\vec x$ is unknown, so we cannot choose $\lambda$ to minimize the exact mean-squared-error. There are a variety of standard techniques for making this choice, with performance that can vary widely depending on the problem being solved, so selection of an appropriate method requires more tests on simulated data. We considered two such methods for $L = D_2$: generalized cross-validation (GCV) and the normalized cumulative periodogram (NCP). We calculate the GCV error using the exact pseudoinverse solution \cite{Vogel:2002ta} and the NCP error by adding together the errors for the real and imaginary parts of the spectra \cite{Hansen:2010vw}. We then minimize these error estimates as a function of $\lambda$ using a 1-dimensional search with the downhill simplex method \cite{Nelder:1965zz}. We also impose the additional restriction $\lambda \geq 5 \times 10^{-11}$ to avoid convergence failures with our SVD implemention that we encountered when performing deconvolutions on noise-free simulated spectra.
The results, also shown in Table~\ref{tab:param-selection}, show that GCV is the best choice for our test problem, with performance nearly matching that of exact selection technique. In contrast, NCP systematically overestimated the noise, as indicated by regularization parameters about four times larger than the exact selection method. However, NCP still offered an improvement in the reduction of the mean-squared-error compared to the naive approach.

\section{Alternative formulation}
\label{sec:alternative_formulation}

There are several obvious extensions or alternatives to the recipe described in section \ref{sec:recipe}. For example, the exact same relations in Eqs.~(\ref{eq:pp-polarization}--\ref{eq:rho2}) hold for a general photon-echo (or non-rephasing) experiment with two distinct pump pulses, except that in this situation the sum in Eq.~\eqref{eq:rho2} to determine the portion of $\rho^{(2)}$ that contributes to the signal should be removed to leave only one of the two phase-matched contributions. Instead, the photon-echo (PE) signal depends on the second-order density matrix given by
\begin{align}
	\rho_\text{PE}^{(2)}(t) = \left(\frac{i}{\hbar}\right)^2 \iint_0^\infty  &dt_2 dt_1  G(t_2) V^{(+)} G(t_1) V^{(-)} \rho_0 \notag \\[-3pt]
	&\times E_2^+ (t - t_2) E_1^- (t - t_2 - t_1).
	\label{eq:rho_2_PE}
\end{align}
However, unlike the case for the state $\rhopp$ that contributes to the pump-probe signal, the excited state portion of $\rho_\text{PE}^{(2)}$ is not necessarily either hermitian or equivalent to the total excited state density matrix. This makes its interpretation less clear but presents no additional technical difficulties for our inversion procedure.

\end{document}